\documentclass[12pt,preprint]{aastex}

\def\pmb#1{\setbox0=\hbox{#1}%
  \kern-.025em\copy0\kern-\wd0
  \kern.05em\copy0\kern-\wd0
  \kern-.025em\raise.0433em\box0 }

\def\bsigma{\pmb{$\sigma$}}

\def\bmu{\pmb{$\mu$}}
\def\bxi{\pmb{$\xi$}}
\def\btheta{\pmb{$\theta$}}

\shorttitle{Radio-wave propagation in non-Gaussian ISM}
\shortauthors{Boldyrev & Gwinn}

\begin{document}
\input psfig.sty

\title{Radio-wave propagation in the non-Gaussian interstellar medium}

\author{Stanislav Boldyrev\altaffilmark{1} and  Carl R. Gwinn\altaffilmark{2}}
\affil{ $^1$Department of Astronomy and Astrophysics, University of Chicago,\\ 
5640 S. Ellis Ave, Chicago, IL 60637; {\sf boldyrev@uchicago.edu}}
\affil{$^2$Physics Department, University of California at Santa Barbara, 
CA 93106; {\sf cgwinn@physics.ucsb.edu}}
\begin{abstract}
Radio waves propagating from distant pulsars 
in the interstellar medium (ISM), are refracted by electron density 
inhomogeneities, so that the intensity of observed pulses fluctuates 
with time. 
The observed pulse shapes are used to diagnose electron-density 
distribution in the ISM. The 
theory relating the observed pulse time-shapes to the electron-density 
correlation function has developed for 30 years,  
however, two puzzles have remained. 
First, observational scaling of pulse broadening with the pulsar distance is 
anomalously strong; 
it is consistent with the standard model only when non-uniform  
statistics of electron fluctuations along the line of sight are assumed. 
Second, the observed pulse shapes are consistent with the standard 
model only when the scattering material is concentrated in a narrow slab  
between the pulsar and the Earth. 

We propose that both paradoxes are resolved at once if one assumes 
stationary and uniform, but {\em non-Gaussian} 
statistics of the electron-density distribution in the interstellar medium. 
Such statistics must be of L\'evy type, 
and the propagating ray should exhibit a L\'evy flight rather than 
the Gaussian random walk implied by the standard model. We propose that 
a natural  realization of such statistics may be provided by the 
interstellar medium 
with random electron-density discontinuities.   
 A L\'evy 
distribution has a divergent second moment, therefore, the standard approach 
based on the electron-density correlation function does not apply. 
We develop a theory of wave propagation in such a non-Gaussian 
random medium, and demonstrate its good agreement with observations.
The qualitative introduction of the approach and the resolution of the 
anomalous-scaling paradox was presented earlier in 
[PRL~{\bf 91}, 131101 (2003); ApJ~{\bf 584}, 791 (2003)].

\end{abstract}

\section{Introduction.}
\label{introduction}
Observations of pulsar signals provide a valuable tool for investigating 
electron density distribution in the interstellar medium. The pulsar 
intrinsic signal is narrow in time, being about few percent of the 
pulsar period. The observed signal (averaged over many periods of 
pulsation) is broad and asymmetric with a sharp rise and a slow decay. 
The pulse broadening is attributed to the random 
refraction the waves experience while propagating in the interstellar 
medium. 

\subsection{Overview.}
\label{overview}

The shapes of the signals, and their scalings with wavelength~$\lambda$ 
 and with the pulsar distance have been investigated 
observationally and analytically for 30 years.  
The standard theory of interstellar 
scintillations (described in detail below) assumes that the propagating 
wave is refracted by random Gaussian gradients, thought to 
be a good approximation due to the central 
limit theorem~\citep{tatarskii}. The symmetric Gaussian distribution is fully 
characterized by its second moment, and the standard theory of 
scintillations aimed at reconstructing this moment from observations. 
The contradictions of this theory with observations 
were noted in the early 1970's, although not many pulsars were 
investigated at that time to make a definitive conclusion.

The signal shape is characterized by its time-width,~$\tau$, 
estimated at the $1/e$ level. 
In 1971, Sutton analyzed scalings of pulse broadening times, $\tau$, 
with the radio wavelength,~$\lambda$, and with 
dispersion measures, $DM$, corresponding to the pulsar's 
distance along the line of sight, $DM=\int dz\, n(z)$. 
Observational scaling for large dispersion measures, 
$DM > 30 \mbox{pc cm}^{-3}$, is close 
to $\tau\propto \lambda^4 DM^4$, while the theory 
gave $\tau\propto \lambda^4 DM^2$. The recent observational results 
are shown in Fig.~(\ref{elbow}). 
To overcome the difficulty 
with the anomalous DM-scaling, 
Sutton suggested that the interstellar turbulence was not statistically 
uniform along the lines of sight, so that the lines of sight for 
more distant pulsars  
intersected regions with 
stronger turbulence. 

The second paradox was encountered by 
\citet{williamson1,williamson2,williamson3}, who compared 
the observed shapes of the pulses with the shapes predicted by 
the standard theory. 
He obtained a surprising result that the model of continuous turbulent 
medium was not consistent with observations. Rather, the best agreement 
was given by the models where all the scattering material was concentrated 
either in a thin screen or in a slab covering only $1/4$ of the line of sight 
between the pulsar and the Earth,  with nearly equal
quality of fit to the data [see the 
discussion and Fig.~4 in \citep{williamson3}]. 

These two assumptions that are necessary to reconcile the standard 
theory with the observations may have physical grounds since the 
interstellar medium is not uniform and denser regions (with stronger 
turbulence) are encountered closer to the galactic center. However, 
in this paper we discuss a simple and perhaps more plausible physical 
explanation for the Sutton and Williamson paradoxes. According to this 
explanation, 
the Sutton and Williamson paradoxes are not the consequence of 
the non-uniform, large-scale galactic distribution of the electron 
density responsible for scattering. Rather, they reflect 
the universal properties of the microscopic structures of the density 
fluctuations.

\subsection{L\'evy flights. Mathematical background.}
\label{levyflights}

In our recent work~\citep{boldyrev1,boldyrev2,boldyrev3} we proposed 
that the Sutton paradox is the evidence of statistically uniform but   
{\em non-Gaussian} electron 
distribution in the interstellar medium. We noted that the 
time-broadened pulses belong to those observational objects 
that depend not on 
the moments of the electron-density distribution function, but on its 
full shape. The standard Gaussian approach did not recognize this 
fact because  
in the Gaussian case, knowledge of the second moment is equivalent to knowing 
the distribution function itself. However, for a general 
non-Gaussian electron-density distribution the second moment may formally  
diverge, and the electron-density correlation function 
may not characterize the pulse shape. 

Physically, this divergence means that the second moment of ray-angle 
deviation is 
dominated not by the bulk of its distribution function, but by 
the extremely far cut-offs at the tails of this function 
[as we will see in the example of Section~\ref{boundaries}]. In contrast, 
the observed signal shape is determined by the bulk of the 
distribution function and is not sensitive to its far-tail cut-offs. 
In other words, 
the electron-density correlation function and the shape of the 
observed pulsar signal provide different, complementary descriptions  
of the electron distribution in the ISM. 

We also note that in the L\'evy model, the pulse shapes may provide 
a unique opportunity to diagnose plasma fluctuations at very small 
scales, $>10^8$~cm. These scales can be smaller than the Coulomb  
mean-free path in HII regions, and can be close to the ion gyro radius. 
Observations of pulse shapes may thus provide diagnostics for magnetized 
plasma fluctuations, which may help to elucidate the role of magnetic 
fields in interstellar turbulence, see also~\citep{armstrong,goldreich}.


In the theory of scintillation, the quantity of interest is the 
transverse electron-density difference, averaged along the line 
of sight, $\Delta n_{\perp}=\int dz\,[n({\bf x}_1, z)-n({\bf x}_2, z)]$ . 
A wave propagating 
in the interstellar medium is scattered (refracted) by density 
inhomogeneities, and the statistics of the refraction angle is 
related to the statistics of density differences 
across the line of sight.   

In the approach proposed in~\citep{boldyrev1,boldyrev2}   
we assumed that 
the distribution function of~$\Delta n_{\perp}$ was stationary and 
uniform, but non-Gaussian, 
and it had a power-law decay as $|\Delta n_{\perp}|\to \infty$. If this 
distribution does not have a second moment, then the sum of many 
density differences along the line of sight 
does not have the Gaussian distribution, i.e. the 
Central Limit Theorem does not hold. Instead, the limiting distribution, 
if it exists, must be the so-called L\'evy distribution, which is common 
in various random systems~\citep{klafter}. By construction, 
such a distribution is stable under convolution, i.e. the 
appropriately rescaled sum of 
two independent variables drawn from the same L\'evy distribution, 
has again the L\'evy distribution. The Gaussian distribution is a 
particular case of this more general stable distribution. 

The Fourier transform (the characteristic function) of 
an isotropic L\'evy probability density function, $P_{\beta}(\bsigma)$, 
has the 
simple form,
\begin{eqnarray}
F(\bmu)=\int\limits_{-\infty}^{\infty}d\bsigma P_{\beta}(\bsigma)
\exp(i\bmu\cdot \bsigma)=\exp(-  |\bmu|^{\beta}),
\label{levy}
\end{eqnarray}
where the L\'evy index $\beta$ is a free parameter, 
$0<\beta \leq 2$. This form is valid for 
an arbitrary dimensionality of the fluctuating vector~$\bsigma $.  
As we explain below, in our case this vector is two-dimensional, 
$\bsigma\propto {\hat {\bf y}}\int dz\,[n(z,{\bf x}_1)-n(z,{\bf x}_2)]$, where 
${\bf x}$ is the coordinate in the plane perpendicular to the line of 
sight, $z$, and ${\hat {\bf y}}$ is 
the unit vector in the direction connecting the two points in this plane, 
${\hat {\bf y}}={\bf y}/|{\bf y}|$, where ${\bf y}={\bf x}_1-{\bf x}_2$.

Equation~(\ref{levy}) 
can be taken as the definition of the {\em isotropic} L\'evy distribution, 
and in the present paper we will consider only isotropic distributions. 
However, it is important to note that a stable distribution need not 
be isotropic [see, e.g.,~\citep{nolan}]. In the simplest case, 
the anisotropic distribution is a copy of the isotropic distribution, 
rescaled along one axis; but far less symmetric L\'evy distributions 
are possible. Non-isotropic stable  
distributions might arise in the case of the magnetized interstellar 
medium, when wave scattering has one or several 
preferred directions. For example, anisotropic scattering may  
be characteristic of interstellar MHD turbulence, 
as described by~\citet{goldreich,lithwick,cho,chandran,scalo}, where 
the preferred direction is given by the large-scale magnetic field. 

The sum of $N$ L\'evy distributed variables scales as 
$\vert \sum^{N}\bsigma \vert\sim N^{1/\beta}$, which becomes diffusion 
in the Gaussian limit $\beta=2$. For $\beta < 2$,  the probability 
distribution function has algebraic tails, 
$P_{\beta}(\bsigma)\sigma \,d\sigma\sim |\bsigma|^{-1-\beta}\,d\sigma$, 
for large $|\bsigma|$, 
and its moments, $\langle |\bsigma|^{\zeta} \rangle$, 
of the order $\zeta \geq \beta$ diverge. 
In \citep{boldyrev1,boldyrev2}, we considered smooth density 
fluctuations, $\sigma\propto |{\bf x}_1-{\bf x}_2|$, and  
the proposed L\'evy statistics of ray-angle deviation  
lead to the pulse-broadening 
scaling $\tau\propto \lambda^4DM^{(2+\beta)/\beta}$.  
This agrees with observations when $\beta\approx 2/3$, and  
provides a natural resolution to the Sutton paradox. 

\subsection{The results of the paper.}
\label{overviewofthispaper}

In this paper we consider scintillations 
produced by a general, non-smooth density fluctuations, $\sigma\propto  
|{\bf x}_1-{\bf x}_2|^{\alpha/2}$; the limit of smooth density 
fluctuations corresponds to~$\alpha=2$.  More precisely, exponent $\alpha$ 
denotes the scaling of the density-difference PDF  
defined in formula~(\ref{levy}), 
i.e. this PDF should have the 
form $P_{\beta}(\sigma, y)\sim P_{\beta}(\sigma/y^{\alpha/2})/y^{\alpha/2}$. 
In the non-smooth case, 
the effective L\'evy index that enters the expressions  
for angle deviations   
is different from $\beta$, and is given by 
$\gamma=\alpha\beta/2$, as is explained 
in formulae~(\ref{levyscreen}) and~(\ref{tau}) below. 

As we noted above, in the L\'evy case  
the electron-density correlation function (or second order 
structure function) is contributed to by the 
far-tail cut-off of the electron-difference distribution function. 
We will denote the correlation-function 
scaling as $\langle \sigma^2\rangle \propto |{\bf y}|^{\delta}$, where 
the scaling exponent, $\delta$, is in general different and 
independent of the L\'evy distribution scaling 
exponents $\alpha$ and $\beta$. The exponent $\delta$ is related to 
the Fourier spectrum of electron density fluctuations, 
$\langle |n(k)|^2 \rangle k^2 \propto k^{-\delta}$.

In the present paper we demonstrate that the proposed L\'evy statistics of 
ray-angle deviations provide a resolution for 
the Williamson paradox as well. For this purpose we, first, 
develop a theory of wave propagation 
in a L\'evy random medium by approximating the medium by a uniform series of 
scattering screens. Then, we provide a general method for constructing 
the pulse-broadening function for an arbitrary L\'evy index~$\gamma$. 
And finally, we compare our results with the observational signals of 
large-dispersion-measure pulsars, recently 
published by~\citet{bhat} and by \citet{ramach}. 
We obtain that the observational shapes 
agree well with the predictions of 
our theory for a uniform medium with the L\'evy index  
$\gamma\sim2/3$ to~$1$, while they are inconsistent with 
the Gaussian theory corresponding to the L\'evy index~$\gamma=2$. 

We also discuss the 
effect of ``over-scattering'' which is inherent for the L\'evy 
scintillations, because the time-signals have long algebraic tails that 
do not decay to zero during pulse periods. This effect is crucial for 
comparing analytical and observational data, 
as we show in Section~\ref{observation}. 

For illustrative purposes, we also  
present a particular model of the density distribution 
that produces the L\'evy statistics 
of scintillations.  
The proposed density distribution is strongly 
spatially intermittent,  and it can be visualized 
as follows. Imagine that the 
electron density is concentrated in separated regions that have 
sharp irregular boundaries.  We make 
the simplest assumption that these boundaries are random, 
similar to randomly folded two-dimensional sheets or shocks.  
If one further assumes that the line of sight intersects the boundary at a 
random angle whose distribution is uniform in all directions, 
such a picture corresponds to the particular case 
of a L\'evy  distribution of $\Delta n$, with $\alpha=2$, $\beta=1$, 
the so-called Cauchy distribution. This distribution is 
rather distinct from the Gaussian one, and is  close to the 
distribution that has been predicted in our theory.  

\subsection{Related models.}
\label{relatedmodels}

The importance of randomly oriented discontinuous objects that can 
be encountered across the line of sight 
has been emphasized in earlier theoretical work on scintillation, 
see e.g., \citep{lambert} and \citep{lazio}. 
As noted   
by \citet{lambert}, sharp density discontinuities may characterize 
stellar wind boundaries, supernova 
shock fronts, boundaries of HII regions at the Str\"omgren radius, etc.
They may also arise from supersonic turbulent motion. \citet{rickett3} proposed 
that similar highly intermittent density structures may be responsible for observed 
low-frequency modulations (fringes) of dynamic pulsar spectra.

The approach  
of \citet{lambert} 
utilized the non-Kolmogorov spectra of discontinuous density fluctuations. 
\citet{lazio} considered wave 
scattering by confined or 
heavily modulated screens (such as disks, filaments, etc.), when the 
statistics of angular deviations were generally assumed Gaussian, with 
parameters varying along the screens. Both papers discussed 
important aspects of non-Kolmogorov and spatially intermittent 
electron-density distribution in the ISM, however, they implied  
the existence of the second moments of the scattering-angle distributions. 
Therefore, in the earlier considerations, the effects that we discuss 
in the present paper could not be discovered. 

As we demonstrate in section~\ref{boundaries}, sharp density discontinuities 
can, in fact, produce a non-Gaussian distribution of the scattering angle, 
whose second moment diverges. In our approach we exploit such intermittent 
density statistics in their full generality. To elucidate the universal role of 
L\'evy distribution, we keep our consideration as simple as possible, 
assuming statistically uniform and isotropic  
scattering screens (although generalizations for the non-isotropic 
and non-uniform cases are possible). The presence of strongly-scattering 
structures is naturally represented in our model by slowly decaying, 
power-law tails of the scattering-angle distributions.  
Our model has a simple physical interpretation and 
provides a practical way of calculating pulse shapes and 
pulse scalings. Most importantly, it naturally resolves both the 
Sutton and the Williamson observational paradoxes, in a manner that is 
simpler and complementary to the standard Gaussian picture. 


The L\'evy-flight model may also be relevant to observations of extreme
scattering events, such as those reported by \citet{fiedler,wolszczan,
stinebring}. Such events
have been investigated theoretically, as by~\citet{rickett3,romani}.  
We note that these
theoretical studies invoke uncommon incidents of scattering much larger
than typical values to explain these events.  In this sense, these
events are consistent with the L\'evy model we develop here, which
includes rare, large events in a statistically-stationary way.
However, we do not yet know whether these phenomena might fit into a
single model for interstellar scattering along with the common
phenomena we discuss in this work.  The answer might possibly depend
on the details of the picture for turbulence in which the L\'evy model
is realized.  Accordingly, we do not discuss the extreme scattering
events in the present work.

\section{Directed-wave propagation in a random medium.}
To address the puzzles mentioned above we need to review the standard 
theory of interstellar scintillations. First, we note that in the 
interstellar 
plasma with typical electron 
density $n\approx 0.03\, \mbox{cm}^{-3}$, the electron plasma frequency 
is $\omega_{pe}=(4\pi n e^2/m_e)^{1/2}\approx 10^4\,\mbox{s}^{-1}$. 
This frequency 
is much smaller than the typical observational frequency of 
$10^8-10^9\,\mbox{Hz}$, and, therefore, the propagating wave scatters 
only by a small angle on the scale of density inhomogeneities. The Fourier 
amplitude of electric field, 
${\bf E}(\omega, {\bf r})=\int {\bf E}(t, {\bf r})\exp(-i\omega t)
\mbox{d}\omega $, 
obeys the following equation:
\begin{eqnarray}
\Delta {\bf E}+\frac{\omega^2}{c^2}{\bf E}-\frac{\omega^2_{pe}}{c^2}{\bf E}=0,
\label{waveeq}
\end{eqnarray}
see, e.g.,~\citep{tatarskii,lee1}. 
We are not interested in polarization effects that are small 
in the considered approximation by a factor $\Omega_e/\omega\sim 10^{-8}$, 
where $\Omega_e=eB/(m_ec)$ is the electron-cyclotron frequency, therefore, 
we consider the scalar wave amplitude,~$E(\omega, {\bf r})$. 

Equation (\ref{waveeq}) can be reduced further, using the so-called 
parabolic approximation~\citep{tatarskii}. 
Assuming that the wave propagates in the 
line-of-sight direction, $z$, we can separate the quickly changing 
phase of the wave from the slowly changing wave amplitude, 
$E(\omega, {\bf r})=\exp(iz\omega/c)\Phi_{\omega}(z,{\bf x})$, 
where ${\bf x}$ is a coordinate perpendicular to~$z$. The equation 
for the wave amplitude $\Phi_{\omega}$ reads:
\begin{eqnarray}
\left[2i\frac{\omega}{c}\frac{\partial}{\partial z}+\Delta_{\perp}-4\pi 
r_0 n(z,{\bf x}) \right]\Phi_{\omega}(z,{\bf x})=0,
\label{paraboliceq}
\end{eqnarray}
where $\Delta_{\perp}$ is a two-dimensional Laplacian in the {\bf x} plane, 
and $r_0=e^2/(m_ec^2)$ is the classical radius of the electron.

Following the approach of \citet{uscinski,williamson4,lee1,lee2,lee3} 
we introduce the two-point function 
$I({\bf r_1},{\bf r_2}, t)=\Phi({\bf r}_1,t)\Phi^*({\bf r}_2,t)$, whose 
Fourier transform with respect to time 
is $I_{\Omega}({\bf r}_1,{\bf r}_2)=1/\sqrt{2\pi}\int d\,\omega 
\Phi_{\omega+\Omega/2}({\bf r}_1)\Phi^*_{\omega-\Omega/2}({\bf r}_2)$. 
For coinciding coordinates, ${\bf r}_1={\bf r}_2$, this function is 
the intensity of the radiation whose variation with time we seek. 
To find this function we need first to solve the equation 
for $V_{\omega, \Omega}\equiv 
\Phi_{\omega+\Omega/2}({\bf r}_1)\Phi^*_{\omega-\Omega/2}({\bf r}_2)$,  
which can be derived from Eq.~(\ref{paraboliceq}). Assuming 
that $\Omega \ll \omega$, we obtain the following  equation,
\begin{eqnarray}
i\frac{\partial V}{\partial z}=
-\frac{1}{2k_1}\frac{\partial^2 V}{\partial {\bf x}^2_1}
+\frac{1}{2k_2}\frac{\partial^2 V}{\partial {\bf x}^2_2} 
+\frac{2\pi r_0}{k}\Delta n V,
\label{vequation}
\end{eqnarray}
where we denote $k=\omega/c$, $\Delta n=n(z, {\bf x}_1)-n(z, {\bf x}_2)$, 
and  $\Delta k\equiv k_1-k_2=\Omega/c$.

Equation (\ref{vequation}) is hard to solve since $n(z, {\bf x})$ is 
an unknown random function. The standard theory uses the fact that    
it takes many refraction events to appreciably deviate the ray 
trajectory, as we discussed in the introduction. Therefore, due to 
the central limit theorem, the ray deviation angle exhibits a Gaussian 
random walk. One can therefore assume that the density fluctuations 
are Gaussian with the specified second-order correlator 
$\langle [n(z_1, {\bf x}_1)-n(z_2, {\bf x}_2)]^2 \rangle 
=2 \kappa({\bf x}_1-{\bf x}_2)\delta(z_1-z_2)$, where short 
correlation length in $z$ direction is the mathematical expression 
of the fact that the ray 
becomes appreciably deviated only when it travels the distance much 
larger than the the density correlation 
length~\citep{tatarskii,lee1,lee2,lee3,rickett1,rickett2}. The 
correlation function $\kappa(y)$ is the two-dimensional 
Fourier transform of the spectrum of the density fluctuations. 
More precisely, if the three-dimensional spectrum of the density 
fluctuations is $D(k)=\langle |n_k|^2\rangle$, 
then $\kappa(y)$ is the Fourier transform of 
$D({\bf k}_{\perp}, k_z=0)$ with respect to ${\bf k}_{\perp}$. 
When the density 
has a power-law spectrum, $D(k)k^2 \propto k^{-\delta}$, with $1<\delta<2$, 
then $\kappa(y)\propto y^{\delta}$. 

As we mentioned in 
the introduction, the statistically uniform Gaussian model fails 
to reproduce the 
observational scaling of the the broadening time, 
$\tau\sim \lambda^4 DM^4$. Numerous 
attempts to reproduce this scaling by using different spectra of 
density fluctuations within this model have not been 
successful either. 
The purely shock-dominated density 
distribution has the spectrum $\langle |n_k|^2\rangle k^2 \propto k^{-2}$, 
while the Kolmogorov turbulence has the 
spectrum~$\langle |n_k|^2\rangle k^2 \propto k^{-5/3}$; the 
difference in the spectral exponents is rather small to have  
a considerable consequence. For various important 
aspects of the standard Gaussian scintillation theory 
we refer the reader to~\citep{goodman,blanford,gwinn,lambert,lithwick}.
In other contexts, the theory of wave propagation 
in Gaussian random media 
was developed in~\cite{saul,jayannavar}.

In the next section we formulate the problem by assuming that the 
refraction occurs in a series of discrete structures, refracting  
`screens', that are statistically identical, independent, and 
placed uniformly along the line of sight.  This setting is 
physically appealing since, as 
we have mentioned in the introduction, 
the refraction in the interstellar medium is consistent with the 
presence of spatially intermittent scattering structures, 
see e.g.,~\citep{lambert,lazio}. 
Moreover, this is the simplest setting when the problem admits 
an exact analytic 
solution that allows us to treat both Gaussian and L\'evy cases on 
the same footing.

\section{Multi-screen model of scintillations.}
\label{multi-screen}

Let us assume that wave scattering occurs in a series of 
thin flat screens uniformly placed along the line of sight. Also, for 
simplicity assume that the incident wave is planar, although an analogous 
consideration can be made for spherical geometry as well.
Each screen gives a contribution to the signal phase; if we 
denote the wave function just 
before the screen by $V^{in}({\bf x}_1, {\bf x}_2)$, then 
the wave function right after the screen 
will be given by $V^{out}({\bf x}_1, {\bf x}_2)=S({\bf x}_1, {\bf x}_2)
V^{in}({\bf x}_1, {\bf x}_2)$. The phase  
function, $S$, can be found from Eq.~(\ref{vequation}), 
\begin{eqnarray}
S({\bf x}_1, {\bf x}_2)=\exp\left(\frac{-2i\pi r_0}{k}
\int\limits_0^{l}dz\,[n(z, {\bf x}_1)-n(z, {\bf x}_2)] \right),
\label{screen}
\end{eqnarray}
where the integration is done over the thickness of the phase screen,~$l$. 
Between the phase screens, the wave propagation is free.
Effecting the change of variables, ${\bf y}={\bf x}_1-{\bf x}_2$, 
and ${\bf x}={\bf x}_1+{\bf x}_2$, 
we rewrite the Eq.~(\ref{vequation}) in this region,
\begin{eqnarray}
i\frac{\partial V}{\partial z}=\frac{\Delta k}{k^2}\left(
\frac{\partial^2 V}{\partial {\bf y}^2}+
\frac{\partial^2 V}{\partial {\bf x}^2}\right)
-\frac{1}{k}\frac{\partial^2 V}{\partial {\bf x}\partial {\bf y}}. 
\label{freeequation}
\end{eqnarray}

In what follows we will be interested in the wave function, $V$, averaged 
over different realizations of the electron density in the phase screens. 
Due to space 
homogeneity, the averaged 
transfer function, ${\bar S}\equiv \langle S({\bf x}_1, {\bf x}_2)\rangle$, 
should depend only on the coordinate 
difference ${\bf y}$. Since refraction affects only the 
${\bf y}$~dependence of the wave function, we can assume that the averaged 
wave function is independent of ${\bf x}$, 
$\langle V(z, {\bf x}_1, {\bf x}_2) \rangle =U(z,{\bf y})$. 
Taking into account the ${\bf x}$~dependence would lead to more 
cumbersome formulas, although it would not qualitatively change the results. 
We therefore model the free propagation between the screens by 
\begin{eqnarray}
i\frac{\partial U}{\partial z}=\frac{\Delta k}{k^2}
\frac{\partial^2 U}{\partial {\bf y}^2}. 
\label{freeequation1}
\end{eqnarray}

To describe the scattering we need to specify the averaged 
phase function,~${\bar S}$. In 
the Gaussian case, one can assume that the screen width, $l$, 
is of the order of 
the characteristic length of the density fluctuations, $l_0$, 
and therefore\footnote{The assumption $l\sim l_0$ is not necessary. The 
only assumption required for our consideration is $l_0\lesssim l$, and the 
final results can be generalized for this case. This generalization 
would however require further assumptions about the scattering 
structures, which at the present level of understanding can hardly 
be justified.},
\begin{eqnarray}
{\bar S}=\exp\left[-\lambda^2 r^2_0l_0\kappa(y) \right],
\label{gaussianscreen}
\end{eqnarray}
where $\lambda=2\pi/k$. Assuming that the main contribution to the 
scattering comes from the scales,~$y$, much smaller than the density 
correlation length, we can 
expand $\kappa(y)\approx \kappa_0(y/l_0)^{\alpha}$, where 
$\kappa_0\sim l_0(\Delta n_0)^2 $, and $\Delta n_0$ is a typical amplitude 
(say, rms value) of density fluctuations. For example, 
if density fluctuations arise due 
to passive advection by the Kolmogorov turbulence, then $\alpha=5/3$. The 
smooth density field corresponds to $\alpha=2$.

The situation is completely different in the L\'evy case, when 
the second-order
moment of the density integral in~(\ref{freeequation}) diverges. 
We however can assume 
that the density difference has some scaling 
form, $\Delta n_{\perp}=\int_0^{l_{0}}[n({\bf x}_1, z)-n({\bf x}_2, z)]dz 
\sim l_0 \Delta n_0 (y/l_0)^{\alpha/2}$, and use the 
L\'evy formula~(\ref{levy}) to write the most general expression 
for the averaged phase function~${\bar S}$,
\begin{eqnarray}
{\bar S}=\exp\left(-|\lambda r_0 l_0\Delta n_0|^{\beta}
|y/l_0|^{\alpha \beta/2} \right). 
\label{levyscreen}
\end{eqnarray}
In the Gaussian limit, $\beta=2$, this formula reduces to the 
previous result (\ref{gaussianscreen}). Similar to the Gaussian case, 
the exponent $\alpha/2$ has the meaning of the  
density-difference scaling with the point separation, however, 
it characterizes not the second moment (that does not exist), but  
the density-difference distribution function itself. 

More precisely, for a chosen point separation ${\bf y}$, 
the projected density difference in the screen function (\ref{screen}), 
\begin{eqnarray}
\Delta \Phi =\lambda r_0
\int_0^{l_0}dz\,[n(z, {\bf x}_1)-n(z, {\bf x}_2)]=\lambda r_0 \Delta n_{\perp}, 
\label{deltaphi}
\end{eqnarray}
should be drawn from a distribution function that depends only on the 
combination $\Delta \Phi/y^{\alpha/2}$. We also note that contrary to the 
Gaussian case, the quantity $\Delta n_0$ entering 
equation~(\ref{levyscreen}), does not correspond to the rms value of 
density fluctuations, since $\langle \Delta n_{\perp}^2 \rangle$ does 
not exist 
in the L\'evy case. Rather, it denotes the typical width of the  
density-difference PDF, say the density difference at which  
this PDF decays by the factor $1/e$ compared to its maximum value. 
On the contrary, the rms value of $\Delta n_{\perp}$ would be given 
by far, non-L\'evy tails of this PDF, and would be much larger 
than~$\Delta n_0$.

Physically, the L\'evy screen function in the form (\ref{levyscreen}) could 
correspond to the 
distribution of the phase integral, $\Delta \Phi$, when the 
integration path intersects a random density discontinuity, say a  
sharp boundary of an ionized region. Such a random 
boundary may naturally arise as a result of turbulent advection  
in an ionized region, and numerical simulations could test this 
intriguing possibility. Our suggestion is based on a  result 
that a smooth, randomly oriented boundary reproduces 
the particular form of the L\'evy screen function (\ref{levyscreen}) 
corresponding to $\beta=1$, $\alpha=2$. We will derive this result 
in the next section. In general, particular values of $\alpha$ 
and $\beta$ depend on the physical realization of scattering screens, 
which is not completely understood and cannot be 
specified in this work.  
For our present purposes, we simply assume the 
general form of the screen function (\ref{levyscreen}) and provide a 
method for calculating the shape and the scaling of the received signal. 

Let us consider a series of such screens placed along the line of sight, $z$, 
perpendicular to it, such that the distance between two 
adjacent screens is $z_0$. Number the screens from the source to the 
observer, and consider the $m$th screen.  Denote as $U_m^{in}(y_m)$ 
the wave function just 
before the wave passes the screen, and $U_{m}^{out}={\bar S}U_{m}^{in}$ this 
function just after the screen. Between the screens $m-1$ and $m$, 
the propagation 
is free, therefore, we obtain from equations 
(\ref{vequation}) and~(\ref{freeequation1}), 
\begin{eqnarray}
U_{m}^{in}(y_{m})=\frac{ik^2}{2\Delta k \pi z_0}
\int \exp\left(\frac{-ik^2({\bf y}_m-{\bf y}_{m-1})^2}{2\Delta k z_0} \right)
{\bar S}(y_{m-1}) U_{m-1}^{in}(y_{m-1})\,d^2y_{m-1}.
\label{uintegral}
\end{eqnarray}
We can write the wave function after the wave has passed $N$ screens,  
by iterating this formula $N$ times, and by using the 
expression for the screen function, (\ref{levyscreen}),
\begin{eqnarray}
& U_{N}^{in}(y_{N})=\left[-i\pi A\Delta k\right]^{-N}
\int \exp\left([iA\Delta k]^{-1}
\sum\limits_{m=1}^{N}(\Delta {\bf y}_m)^2 
-B^{\alpha \beta/2}\sum\limits_{m=1}^{N-1}
\vert y_{m} \vert^{\alpha\beta/2} \right)\times \nonumber \\
& U_{0}(y_{0})\, 
d^2y_{N-1}, \dots, d^2y_1\,d^2y_{0}
\label{unscreens}
\end{eqnarray}
where we introduced the short-hand notation: 
$A=2z_0/k^2$, $B=|\lambda r_0 l_0 \Delta n_0|^{2/\alpha}/l_0 $, 
and $\Delta {\bf y}_m={\bf y}_m-{\bf y}_{m-1}$. 

To do the integrals, 
we need first to make a simple transformation of 
formula~(\ref{unscreens}). We will substitute the following 
identity,
\begin{eqnarray}
\exp\left[(\Delta {\bf y}_m)^2/(iA\Delta k)\right]=
-i\pi A\Delta k \int \exp\left(i\bxi_m\cdot 
\Delta {\bf y}_m-i A \Delta k \xi_m^2\right)d^2\xi_m, 
\label{identity}
\end{eqnarray}
for each $\Delta y_m$, into formula~(\ref{unscreens}). Our ultimate goal 
is to find the time dependence of the pulse intensity at the $N$th screen, 
i.e., $I(t)=\int U_N(y_N=0, \Delta k)\exp(i\Delta k\, c\, t)d\Delta k\, c$. We 
therefore change the order of integration and do the $\Delta k$-integral 
first. As the result we get:
\begin{eqnarray}
&I(t)=\int \exp\left(-B^{\alpha\beta/2}
\sum\limits_{m=1}^{N-1}|y_{m}|^{\alpha\beta/2}
-i\sum\limits_{m=1}^N\bxi_m\cdot \Delta {\bf y}_m \right)U_0(y_0)\times \nonumber \\
&\delta\left(t-A\sum\limits_{m=1}^N\xi^2_m/c\right)\,
d^2\xi_N\dots d^2\xi_1\,d^2y_{N-1}\dots d^2y_0.
\label{intensity}
\end{eqnarray}
Introducing the non-dimensional variables ${\tilde y}_m=y_mB$ and 
${\tilde \xi}_m=\xi_m/B$, we rewrite formula (\ref{intensity}) as follows:
\begin{eqnarray}
&I(t)=\int \exp\left(
-\sum\limits_{m=1}^{N-1}|{\tilde y}_{m}|^{\alpha\beta/2}
-i\sum\limits_{m=1}^N {\tilde {\bxi}}_m\cdot \Delta {\tilde {\bf y}}_m \right)
U_0({\tilde y}_0/B)\times \nonumber \\
&\delta\left(t-\frac{AB^2}{c}\sum\limits_{m=1}^N{\tilde \xi}^2_m\right)\,
d^2{\tilde \xi}_N\dots d^2{\tilde \xi}_1\,d^2{\tilde y}_{N-1}\dots 
d^2{\tilde y}_0.
\label{intensity-non-dim}
\end{eqnarray}
In the rest of this section  
we will use only the non-dimensional variables, and will omit the 
tilde signs.

Now we are ready to do the $y$-part of the integral. For this we make a 
simple rearrangement in the sum 
$\xi_1(y_1-y_0)+\dots +\xi_N(y_N-y_{N-1})\equiv 
-y_0\xi_1+y_N\xi_N+y_1(\xi_1-\xi_2)+\dots +y_{N-1}(\xi_{N-1}-\xi_{N})$. 
Recalling that $y_N$ should be 
set to zero, we are left with 
$\sum_{m=1}^N\bxi_m\cdot \Delta {\bf y}_m =-y_0\xi_1-\sum_{m=1}^{N-1}
{\bf y}_m \cdot \Delta \bxi_{m+1}$, where $\Delta \bxi_{m}=\bxi_m-\bxi_{m-1}$.
To complete the $y$-integration, we simply use the L\'evy 
formula (\ref{levy}), and get:
\begin{eqnarray}
I(t)=\int\delta\left(t-\frac{A B^{2}}{c}
\sum\limits_{m=2}^N\xi^2_m\right) 
P_{\alpha\beta/2}(\Delta \bxi_N)\dots P_{\alpha\beta/2}(\Delta \bxi_2)
{\tilde U}_0(\xi_1 B)d^2\xi_N\dots d^2\xi_1, 
\label{levyintensity}
\end{eqnarray}
where ${\tilde U}_0(\xi)$ is the Fourier transform of $U_0(y)$, $P$ is the 
L\'evy distribution function introduced in~(\ref{levy}). 
The last step is 
to note that $d^2\xi_N\dots d^2\xi_2=d^2\Delta \xi_N\dots d^2\Delta \xi_2$, 
which allows us to give a simple interpretation to 
formula~(\ref{levyintensity}). 

Before discussing this interpretation, we note the quantum mechanical 
analogy of our approach. If we could assume  
that $B^{\alpha \beta /2}\propto z_0$, and $z_0\propto 1/N$ 
in the limit when the number of screens increases, $N\to \infty$, but 
the line-of-sight length is constant (which would be natural for 
the Gaussian case), the integral that we calculated in (\ref{unscreens}) 
would formally  become the Feynman-Kac path 
integral for the solution of the Schr\"odinger equation with the 
potential~$\propto i|{\bf y}|^{\alpha \beta/2}$. The transition 
from (\ref{unscreens}) to  (\ref{levyintensity}) given by 
formula~(\ref{identity}) is the transition from 
the coordinate to the momentum representation of this integral. 

Naturally, in the Gaussian case one can find the shape of~$I(t)$ 
either by doing the multiple integral in~(\ref{levyintensity}) or 
by solving the corresponding Schr\"odinger equation. 
The first approach was essentially adapted in \citep{uscinski,williamson4}, 
the second one in \citep{lee2}. In our L\'evy model we cannot assume 
that $B^{\alpha \beta /2}\sim 1/N$, therefore, we have to work with 
the general expressions~(\ref{unscreens})-(\ref{intensity}). 
Although the general integral (\ref{unscreens}) is quite complicated,  
the quantity of interest, $I(t)$, which is calculated with the 
aid of this integral, has a quite simple meaning, and its probabilistic 
interpretation~(\ref{levyintensity}) can be easily understood.  

The $\delta$-function in formula (\ref{levyintensity}) means that the signal 
intensity $I(t)$ is the probability density function of the time delay, 
$\tau=AB^2\sum_{m=1}^{N} \xi^2_m/c$.  
The variable $\bxi_m$ is proportional to the deviation angle of 
the ray path from the $z$-axis, 
$\bxi_m\propto \btheta_m$. [This is clear from~(\ref{identity}) since 
$\bxi_m\propto\Delta {\bf y}_m/z_0$, and 
this will also be the case in the geometric-optics analysis 
of Section~\ref{boundaries}.]
This angle is the sum of elementary 
angle deviations  caused by each phase 
screen, viz $\bxi_m=\bxi_1+\sum_{i=2}^m\Delta \bxi_i$. 
The propagation time between 
two neighboring screens exceeds the propagation time 
along the $z$-axis by the amount $\Delta \tau=({z_0}/{c})[1-\cos\theta_m] 
\propto \xi_m^2$, and to get the total time delay, $\tau$, we need 
to sum up these individual time delays. 

Formula~(\ref{levyintensity}) 
teaches us that each angle increment, $\Delta \bxi_s$, has a L\'evy 
distribution with 
the index $\gamma=\alpha\beta/2$, which provides a practical way of 
constructing the pulse broadening function~$I(t)$ without 
doing the multiple integrals 
in~(\ref{levyintensity}). This is a Monte-Carlo-type method for 
multiple integration.   
Indeed, to find the effect caused by 
the interstellar medium, we can 
neglect the intrinsic pulse shape, i.e., we
can assume that the initial angle of the planar wave is 
zero, $\bxi_1=0$. Then the function 
$I(t)$ is the probability density function of the variable
\begin{eqnarray}
\tau=\frac{z_0 \left[r_0 l_0\Delta n_0\right]^{4/\alpha}}{2l_0^2\pi^2c}\lambda^{2+4/\alpha}
\sum\limits_{m=2}^N\left[\sum\limits_{s=2}^m\Delta \bxi_s \right]^2,
\label{tau}
\end{eqnarray}
where all the variables $\Delta \bxi_s$ are distributed independently and 
identically according to the L\'evy law with the 
index $\gamma=\alpha\beta/2$. The same expression was obtained in 
our previous work for the case $\alpha=2$, although without a detailed 
derivation~\citep{boldyrev1,boldyrev2,boldyrev3}. 
We remind that the scaling exponent $\alpha$ corresponds to the 
spatial scaling of the density field, as is defined 
in Eq.~(\ref{levyscreen}). The Gaussian case corresponding 
to $\alpha=2$, $\beta=2$ 
was originally considered by \citet{williamson1}. The Gaussian case 
corresponding to the Kolmogorov density fluctuations, 
$\alpha=5/3$, $\beta=2$ was considered by~\citet{lee3}; however, the central 
part of the signal $I(t)$ predicted by their model did not 
differ much from the Williamson case. 

We do not know the analytic expression for the probability density 
function of~$\tau$. However, it can be calculated numerically, by 
the method described in \citep{chambers,nolan}. The generated 
pulse shapes for different numbers of screens,~$N$, seem to be 
universal when appropriately rescaled. We will construct  
such shapes and will compare them with observations in 
Section~\ref{observation}.

The scaling of the time delay $\tau$, i.e. the approximate scaling of 
its probability density function, can be found from the following 
consideration. Using the expression for the scaling of the sum 
of L\'evy distributed 
variables, $\vert\sum^N\Delta \bxi\vert \sim N^{1/\gamma}$ (recall that 
the Central Limit Theorem does not hold), we derive 
$\tau\propto \lambda^{2+4/\alpha} N^{(2+\gamma)/\gamma}$, 
where the number of screens is proportional to the distance 
to the pulsar, $N\propto DM$. The observational 
scaling, $\tau\propto \lambda^4 DM^4$, 
is reproduced for $\alpha\approx 2$, 
and $\beta \approx 2/3$, which shows that ray-deviating density 
fluctuations do not  
have a Gaussian distribution but rather a L\'evy distribution 
with the index $ \beta\approx2/3$. This is the important 
result of the L\'evy model of scintillations, and it resolves 
the Sutton paradox.

In the Section~{\ref{observation}}, we will demonstrate that the Williamson 
paradox is resolved naturally in the L\'evy picture of scintillations as well. 
But before that, in the next section we would like to present a 
physical model of scintillations that can be easily visualized and that 
demonstrates the essence of our approach.

\section{Scintillations caused by random density discontinuities}
\label{boundaries}
The intensity function $I(t)$ can be calculated for the  
phase-screen function (\ref{levyscreen}), once 
the parameters $\alpha$ and $\beta$ have been specified from the physics of 
the problem. As an important example, let us consider a particular 
case of L\'evy phase screens, which we shall call Cauchy screens, when  
the values of these parameters can be derived directly. Let us assume 
that the interstellar medium is filled with separated ionized regions 
such that all of them have the same electron-density contrast 
with the ambient medium,~$\Delta n_0$. We also assume 
that these regions are randomly shaped and have sharp boundaries. 

A propagating ray is then refracted by these random boundaries due to 
the Snell's law, ${\eta}_1\sin(\theta_1)={\eta}_2\sin(\theta_2)$, 
where $\theta$ 
is the angle between the ray and the normal to the boundary, 
and ${\eta}=({1-\omega_{pe}^2/\omega^2})^{1/2}\approx 
1-\omega_{pe}^2/(2\omega^2)=1-(2\pi)^{-1}\lambda^2 r_0 n$  
is the refraction index. As we explained in the introduction, the 
interstellar parameters ensure that $\omega_{pe}^2/\omega^2\ll 1$, and each 
refraction event results in a very small angle deviation, 
$\delta\theta=\theta_1-\theta_2$. Expanding $\sin(\theta_1)
=\sin(\theta_2+\delta\theta)$ through the first order in $\delta \theta$, 
we get from the Snell's law
\begin{eqnarray}  
\delta \theta=(2\pi)^{-1}\lambda^2 r_0 \Delta n_0 \tan (\theta),   
\label{theta}
\end{eqnarray}
where $\Delta n_0=n_1-n_2$.
The distribution of $\tan(\theta)$ can be found assuming (somewhat 
artificially) that the 
boundary normal is uniformly distributed over all directions in a 
half-space, at the point where it intersects the line of sight. 
Denote $\bsigma=\tan(\theta){\bf n}$, where~${\bf n}$ is a unit vector 
indicating the direction of the refraction in the plane perpendicular 
to the ray propagation. Then the distribution of~${\bsigma}$ is 
given in the polar coordinates by:
\begin{eqnarray}
P(\bsigma)\,d\sigma\,d\phi=\frac{1}{2\pi}
\frac{\sigma}{(1+\sigma^2)^{3/2}}\,d\sigma\,d\phi.
\label{cauchy}
\end{eqnarray}
Quite remarkably, this distribution is the two-dimensional 
L\'evy distribution with the 
index $\beta=1$, also known as the Cauchy distribution. Its Fourier 
transform follows from formula~(\ref{levy}), 
 $F(\bmu)=\exp(-|\bmu|)$, we leave the derivation of this 
result to the reader as an instructive exercise.

To apply this result we need to calculate the screen function~(\ref{screen}).
It can be  done if we note that for the plane-like density 
discontinuity, the screen phase~(\ref{deltaphi}) is easily calculated, 
$\Delta \Phi=\lambda r_0\int dz\, [n(z,{\bf x}_1)-n(z,{\bf x}_2)]=
\lambda r_0 \Delta n_0  \,
(\bsigma\cdot {\bf y})$. Since $\bsigma$ has the Cauchy 
distribution~(\ref{cauchy}), the averaged 
screen function can be found with the aid of formula~(\ref{levy}), 
\begin{eqnarray}
{\bar S}=\exp\left(- \lambda r_0 \Delta n_0 |{\bf y}| \right),
\end{eqnarray}
and we recover formula (\ref{levyscreen}) with $l=l_0$, and 
with $\alpha=2$, $\beta=1$; 
the parameters are close to those that 
we have proposed for the interstellar 
scintillation on the observational grounds, i.e.,  $\alpha\approx 2$, 
$\beta\approx 2/3$. We expect that the observational 
scaling $\beta\approx 2/3$ would correspond 
to more complicated, realistic structure of the boundaries, 
possibly, with the fractal dimension larger than~2.  
Interstellar turbulence can indeed corrugate two-dimensional structures 
to make them have higher-than-two fractal dimensions, as is seen in numerical 
results and in observations of molecular 
clouds~\citep{boldyrev4,boldyrev5,norman,elmegreen}; see also 
\citep{sreenivasan,constantin,kraichnan}. We expect this effect to 
have implications for interstellar scintillations, and we plan to 
consider it in the future work.

An attentive reader has probably noticed that Eq.~(\ref{theta}) for the 
angle deviation cannot be valid for the  
arbitrarily large $\tan(\theta)$. Indeed, the refraction 
angle $\theta_2$ cannot exceed the critical angle, given by 
$\sin(\theta_c)={\eta}_1/{\eta}_2$, which is obtained when 
$\theta_1=\pi/2$, where we assume ${\eta}_2>{\eta}_1$. 
To obtain the general 
formula for the angle deviation we need to expand  
$\sin(\theta_1)=\sin(\theta_2+\delta \theta)$ in the Snell's law 
up to the second order in $\delta \theta$, which gives  
\begin{eqnarray}
\delta \theta =\tan^{-1}(\theta)
\left[\sqrt{1+\pi^{-1}\lambda^2 r_0 \Delta n_0\tan^2(\theta)}
-1\right].
\end{eqnarray}
The distribution of $\delta\theta$ thus has a cut-off 
at $\delta\theta_c=(\lambda^2 r_0 \Delta n_0/\pi)^{1/2}$. For example, if the 
electron density fluctuations are $\Delta n_0\sim 10^{-2}\,\mbox{cm}^{-3}$, 
we have $(2\pi)^{-1}\lambda^2 r_0 \Delta n_0\sim 10^{-12}\lll 1$, 
which implies that formula~(\ref{theta}) holds in the broad range 
of scales. Indeed,  
the distribution of $\sigma=2 \pi \delta\theta /(\lambda^2 r_0 \Delta n_0)$ 
follows the Cauchy law~(\ref{cauchy}), therefore the 
distribution function of~$\delta \theta$ is strongly peaked at 
small~$\delta \theta$. This function has a maximum at  
$\delta \theta\sim (2\pi)^{-1} \lambda^2 r_0 \Delta n_0 
\sim 10^{-12}$, and a long asymptotic power-law tail, 
$P(\delta\theta)\propto 1/(\delta \theta)^2$, that spans about six  
orders of magnitude before it reaches the  
cut-off at $\delta \theta_c \sim 10^{-6}$. 

This consideration provides an illustration to the main 
idea of our approach.  The second moment of the distribution 
of~$\delta \theta$ depends not on the shape of the distribution 
function~(\ref{cauchy}), but on the cut-off value of its  
asymptotic power-law tail,~$\delta \theta_c$.
On the contrary, the shape of the pulsar signal is 
determined by the full shape of the  
distribution function~(\ref{cauchy}), 
and is practically independent of the large cut-off value~$\delta \theta_c$. 


\section{Comparison with observations.}
\label{observation}

In Fig.~(\ref{levy_cauchy}) we present the pulse-broadening function, 
$I(t)=\langle \delta(t-\tau) \rangle$ which is the distribution function 
of the time delay, $\tau$, given by formula~(\ref{tau}),  
for the standard Gaussian model, $\alpha=2, \beta=2$, 
for the  L\'evy-Cauchy model, $\alpha=2, \beta=1$, and for the L\'evy 
model with  $\alpha=2$, $\beta=2/3$. Note that, in comparison with the 
pulse through a Gaussian medium, the
L\'evy models predict relatively greater intensity at short delays from the
arrival time for the original pulse, as well as power-law tails at long
delays. These tails result from rare, large occurrences of scattering.

We generated the two-dimensional isotropic L\'evy distribution of
$\Delta\bxi_s$ using the methods of~\citet{feller,chambers,nolan}.  
Let ${\bf X}=(X_1, X_2)$ be a Gaussian random vector, i.e. 
both components $X_1$ 
and $X_2$ are independent, identically distributed, one-dimensional 
Gaussian variables. Let $Y$ be 
a completely skewed (i.e. positive), independent L\'evy distributed 
variable corresponding to the L\'evy index $\gamma/2$.  This variable 
can be numerically 
generated using the method by~\citet{chambers}, see also~\citep{nolan}. 
Then the variable $\Delta\bxi= {\bf X}\sqrt{Y}$ has an isotropic 
two-dimensional L\'evy distribution with the index~$\gamma$; the proof 
can be found, e.g., in~\citep{feller}. To generate an
anisotropic L\'evy distribution with uniform index $\gamma$ (as may be
relevant for the magnetized interstellar medium) one may choose an
anisotropic Gaussian vector~{\bf X}, and then proceed as above. In our present
work we consider only isotropic distributions.

We compared the predicted pulse shapes with the data recently published 
in~\citep{bhat} and in~\citep{ramach}.  
Out of the 76 pulsars analyzed by \citet{bhat},
we considered those with broad signals, in order to minimize the 
intrinsic-pulse effects. Good examples are provided by the four pulsars,   
P1849+0127 at 430 MHz (DM=214.4), J1852+0031 at 1175 MHz  (DM=680.0),  
J1905+0709 at 430 MHz  (DM=269.0), 
and P1916+0844 at 430 MHz (DM=339.0), where higher-frequency 
observations indicate that the intrinsic signal may be narrow. The 
dispersion measures are given in the units of $\mbox{pc}\cdot\mbox{cm}^{-3}$.
Shapes of all of the pulses have a characteristic sharp rise and a narrow, 
pointed apex,  which is inconsistent with the Gaussian model, 
in exact agreement with the observations by~\citet{williamson3} 
[see Fig.~(4) in Williamson's paper]. On the contrary, our L\'evy model 
provides a good fit to such shapes. 
For another similar comparison we used the pulse shape 
of J1848-0123 (DM=159.1) observed at 327 MHz by~\citet{ramach}; 
and we present  
this comparison in Figs.~\ref{shapes2},\ref{alpack} as the illustrative example.

When making the comparisons, we tried to match the central parts  
of the observational and analytical curves. However, this  
was possible only when we shifted the zero level of the analytical 
curves down, as is shown by the thin solid lines in 
Fig.~(\ref{shapes2}).  
This may be natural for ``over-scattered'' profiles, i.e., when 
the signal has a long tail that does not approach zero during the 
pulse period.  This is exactly the case for our L\'evy model. 
Besides, other observational effects, including noise from the sky
and the telescope itself, usually prevent one from determining the
baseline unambiguously.

We note that for a precise comparison, we must convolve the calculated
pulse-broadening function with the narrow intrinsic shape of the
pulse, and with the response function of the receiving 
system~\cite{ramach,bhat2}.
This will lead to slight broadening of the predicted pulse
profile, making the Gaussian model even less consistent with the
observations.

The effects of over-scattering on baseline have not been considered 
observationally. This effect is important for comparison of our
theory with observations, and we plan to investigate it in future work.
Interferometry can distinguish between sky and telescope noise, and 
the overscattered tail of the pulse; comparison of the 
correlated flux density at the peak of the pulse, and between pulses,
thus provides a measure of the parameter $\gamma$ in 
Figures~\ref{levy_cauchy} and~\ref{compare}. 

We are proceeding with additional observational tests that 
can distinguish among models for scattering.
As Figures~\ref{levy_cauchy},\ref{compare} suggest, different 
values of $\gamma$
yield different delays between the arrival time for the unscattered
pulse and the peak of the pulse, as well as different behavior in the tail.
Observations at short wavelengths
yield the arrival time for the unscattered pulse,
with corrections for dispersion in the uniform interstellar plasma.
Thus, multi-frequency observations of pulse broadening,
with timing information, can distinguish among the different models.

Shapes of scattered images provide another possible observational test
of our theory.  As Equation~(\ref{levyscreen}) suggests, 
the distribution of
deflections of waves, at a given point, is drawn from a L\'evy
distribution. Because the L\'evy distribution is a stable attractor,
iteration of the process yields a L\'evy distribution of directions, at
any distance from the source.  Compared with a Gaussian distribution
of similar width,
a L\'evy distribution has excesses at both large and small angles:
it produces large deflections more commonly, 
and compensates with a large population with small deflection.
Thus, the scattered image for a L\'evy flight should have a halo
at large angles, and a cusp at small angles,compared with a Gaussian image.
Interestingly, observations of some heavily-scattered extragalactic
sources show such core-halo excesses~\citep{desai}.
Intrinsic structure of the source could possibly explain the observed 
halo structure; but cannot explain the relatively less-scattered cusp.



\section{Discussions and conclusions.}
\label{conclusions}

In the present paper, we have investigated the observational pulse
shape. Another object directly related to the electron distribution 
function is the 
visibility function, $\langle \Phi({\bf x}_1, t)\Phi^*({\bf x}_2,t)\rangle$,  
measured at the locations ${\bf x}_1$ and ${\bf x}_2$ at the Earth. However, 
the electron fluctuations probed by this function are limited to  
small scales, $|{\bf x}_1-{\bf x}_2|\lesssim 10^8$cm, which can be 
comparable to the inner scale of the turbulence~\citep{gwinn2,kaspi,gupta} 
where 
the statistics of the ISM are different from the statistics responsible 
for pulse broadening. The visibility 
function can be easily constructed in our multi-screen model by 
solving Eqs.~(\ref{vequation},\ref{freeequation}) for $\Delta k=0$, 
however, we leave 
its analysis for future communication. Here, we  
estimate what restrictions would be imposed on our theory by an  
inner scale of the order $y_{in}\sim 10^{8} \mbox{cm}$. 

Due to the formulas of Section~\ref{multi-screen}, large-angle refraction 
is provided by small-scale fluctuations of electron density.  In the 
case of L\'evy statistics, a sum of random 
variables can be dominated by a single term, 
therefore, from formula~(\ref{tau}) 
we can estimate $(\Delta {\tilde {\xi}}_{max})^2\sim 
2\pi \tau c/(z_0 N \lambda^2 B^2)$. 
From formula~(\ref{intensity-non-dim}), the electron-density 
fluctuations producing such an angle deviation 
should be strong at the scale $y_{min}\sim 2\pi/(B \Delta {\tilde {\xi}}_{max})
\sim [2\lambda^2 Nz_0/(\tau c)]^{1/2}$, or 
\begin{eqnarray}
y_{min}\sim\sqrt{\frac{2 c DM}{ \langle n \rangle \tau \nu^2}},
\label{yin}
\end{eqnarray}
where we introduced the dispersion measure $DM=\langle n \rangle N z_0$, 
and the signal frequency, $\nu=c/\lambda$. Quite conveniently, 
formula (\ref{yin}) includes only the quantities 
averaged along the line of sight, 
but not the particular parameters $l$, $l_0$, $\Delta n_0$, 
and $\alpha$ of the scattering screens. 

As an example, let us 
consider the parameters for the large-dispersion-measure 
pulsar J1852+0031 (DM=680.0) observed at 1175 MHz in \citep{bhat}.   
Substituting 
$\nu\sim 10^9\,\mbox{s}^{-1}$, $DM\sim 2\cdot 10^{21} \,\mbox{cm}^{-2}$, 
$\tau \sim 0.5\,\mbox{s}$, and assuming that 
$\langle n \rangle \sim 3\cdot 10^{-2}\,\mbox{cm}^{-3}$, 
we obtain $y_{min}\sim 10^8\,\mbox{cm}$. 
 Therefore, the 
condition $y_{min}\sim y_{in}$ is marginally satisfied. 
Due to the scaling $\tau\propto \lambda^4 DM^4$, 
smaller-dispersion-measure pulsars satisfy this condition 
even better. For the cases when this 
condition is violated, a more appropriate description will be given 
by ``truncated'' L\'evy distributions, see e.g.,~\citep{nakao}.


In conclusion we would like to mention the  
observational correlation functions that are expressed through the 
higher-than-second moments of the 
wave amplitude, $\Phi({\bf x},t)$. These observational quantities 
require a special consideration that  
at present is not available to us.  The examples of such quantities  
are given by the frequency correlation function that is used to 
characterize pulse broadening for $DM<100 \,\mbox{pc cm}^{-3}$, see, 
e.g., \citep{gwinn}, or by 
the intensity correlation functions that are used to characterize 
the density statistics on scales larger than the Earth size \citep{shishov}. 
The scaling of such functions may be easily found in the Gaussian 
theory when these functions can be related to the second-order moment,  
$\langle (\Delta n)^2 \rangle$.  
However, the meaning of these 
correlation functions 
in the L\'evy case has yet to be determined. For instance, it seems 
possible that the deviation of the pulse-width scaling from  
$\tau\propto DM^4$ 
for the lower part of the elbow diagram, 
 in  Fig.~(\ref{elbow}),  
is due to the different method of reconstructing $\tau$ in this region. 
Indeed, 
below $DM\sim 100\mbox{pc cm}^{-3}$,  
$\tau$ is not measured directly, but rather is defined as the inverse 
frequency decorrelation 
bandwidth~\citep{sutton,rickett1,rickett2,ramach,gwinn,bhat}. 

To summarize, in the present paper we proposed a multi-screen 
model of the uniform, 
non-Gaussian interstellar medium, and we found its analytic solution. 
We demonstrate that this model explains both the shapes and 
the scalings of the observational pulse 
profiles, and it is free of the Sutton and Williamson paradoxes that 
could hardly be resolved in  the standard Gaussian theory of 
scintillations. Also, we provided a physical model of density 
distribution having random discontinuities,  
which produces L\'evy scintillations and which may serve as a 
good approximation for the analysis of the observational data. 
We have demonstrated that the observational pulse shapes depend on the
entire shape of the distribution of electron density, rather than
simply on the second moment. Pulse profiles may therefore serve 
as a valuable 
tool for reconstructing this function from observations, 
and we have developed  
the corresponding method in the present paper.

\section{Acknowledgement}
We are very grateful to 
Ramachandran Rajagopalan for 
allowing us to use the observational results published in \citep{ramach}. 
We would like to thank John Nolan 
for his valuable advice on the method of numerical generation 
of symmetric stable distributions, and Fausto Cattaneo, 
Arieh K\"onigl, and Robert Rosner for important discussions. S.B. 
would like to thank the Aspen Center for Physics where a part of 
this work was done. The work of S.B. was supported by 
the NSF Center for magnetic self-organization in astrophysical and 
laboratory plasmas, at the University of Chicago.

\clearpage
{
\begin{figure} [tbp]
\centerline{\psfig{file=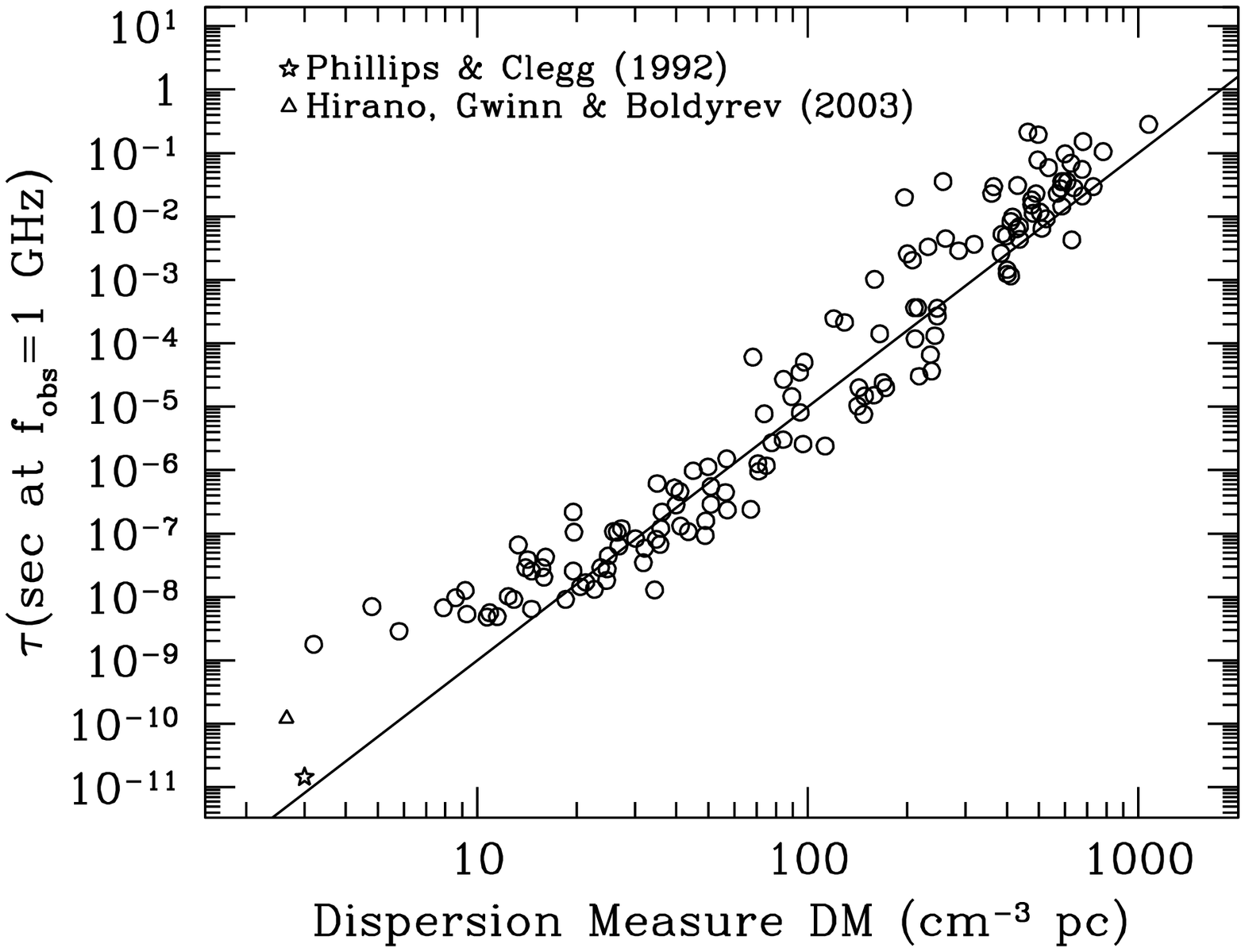,width=6.in,angle=0}}
\vskip3mm
\caption{Pulse temporal broadening as a function of
the dispersion measure, $\mbox{DM}=\int_0^dn(z)\,\mbox{d}z$, which is a
measure of the distance to the pulsar. Except as noted, data were taken 
from (Phillips \& Clegg 1992). The solid
line has slope~4, which contradicts the stangard Gaussian 
model of scintillations predicting slope 2 (the Sutton paradox), 
and which agrees with the L\'evy model with~$\beta=2/3$.
}
\label{elbow}
\end{figure}
}

\newpage
{
\begin{figure} [tbp]
\centerline{\psfig{file=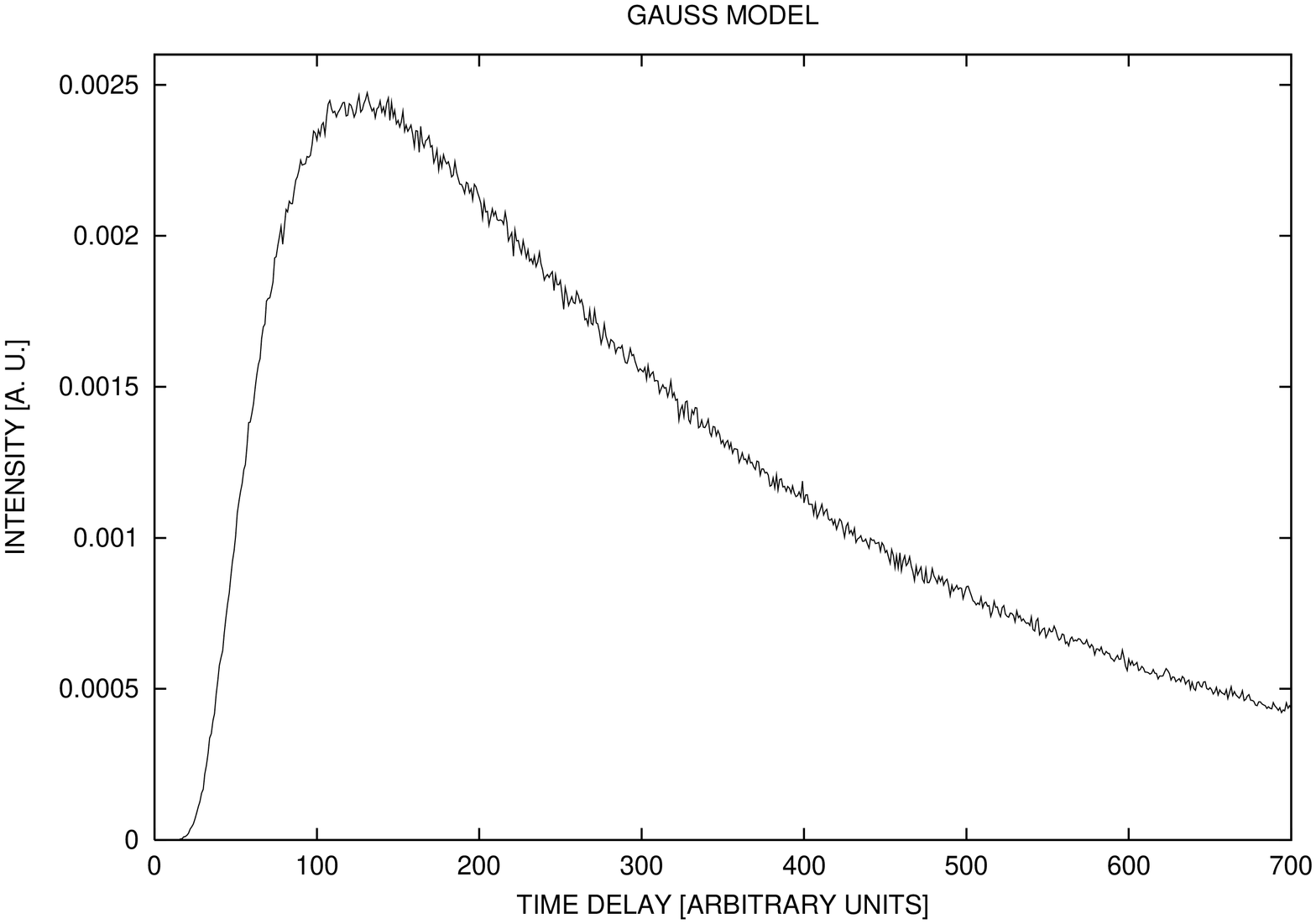,width=3.5in,angle=-90}}
\vskip3mm
\centerline{\psfig{file=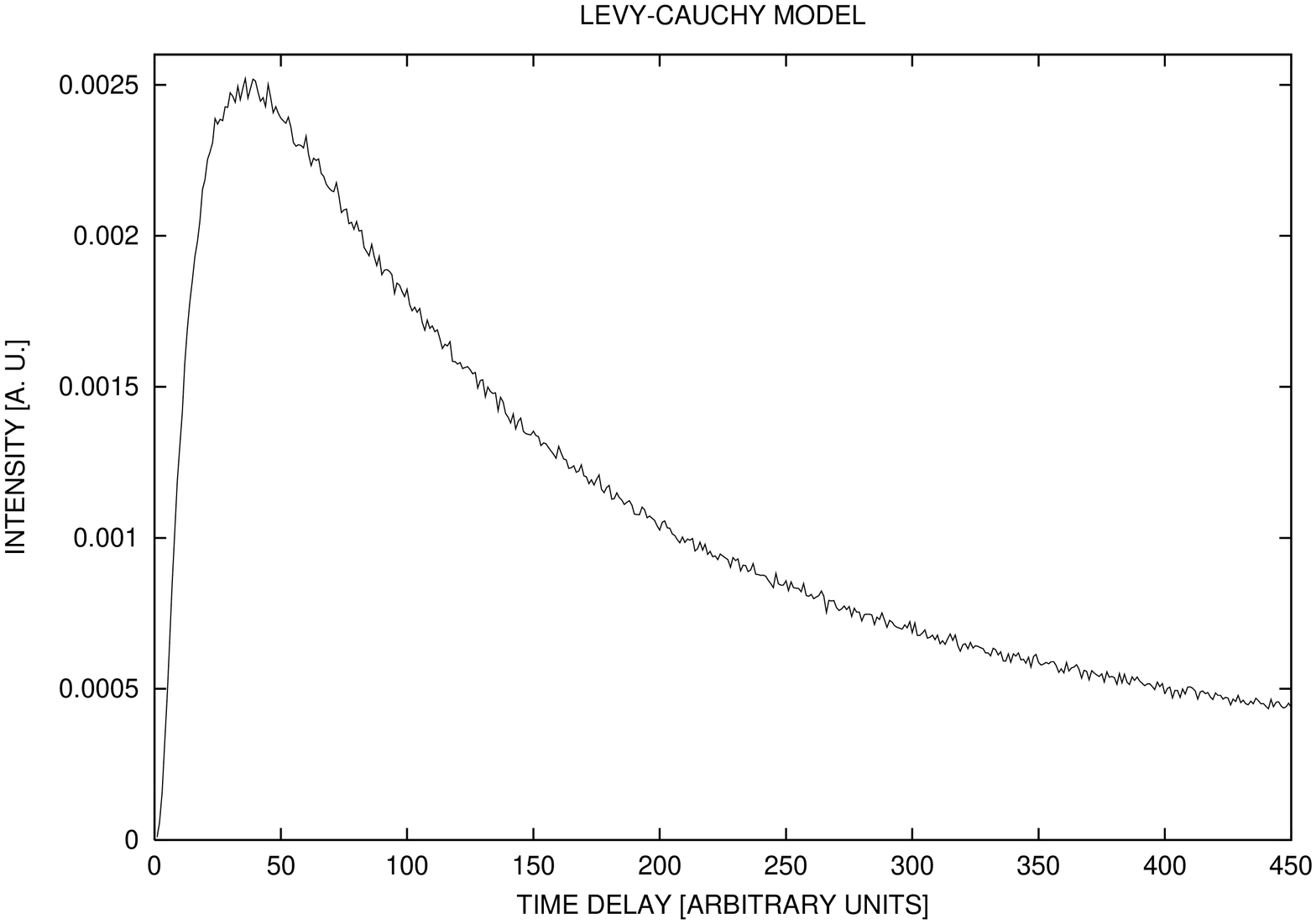,width=3.5in,angle=-90}}
\vskip3mm
\centerline{\psfig{file=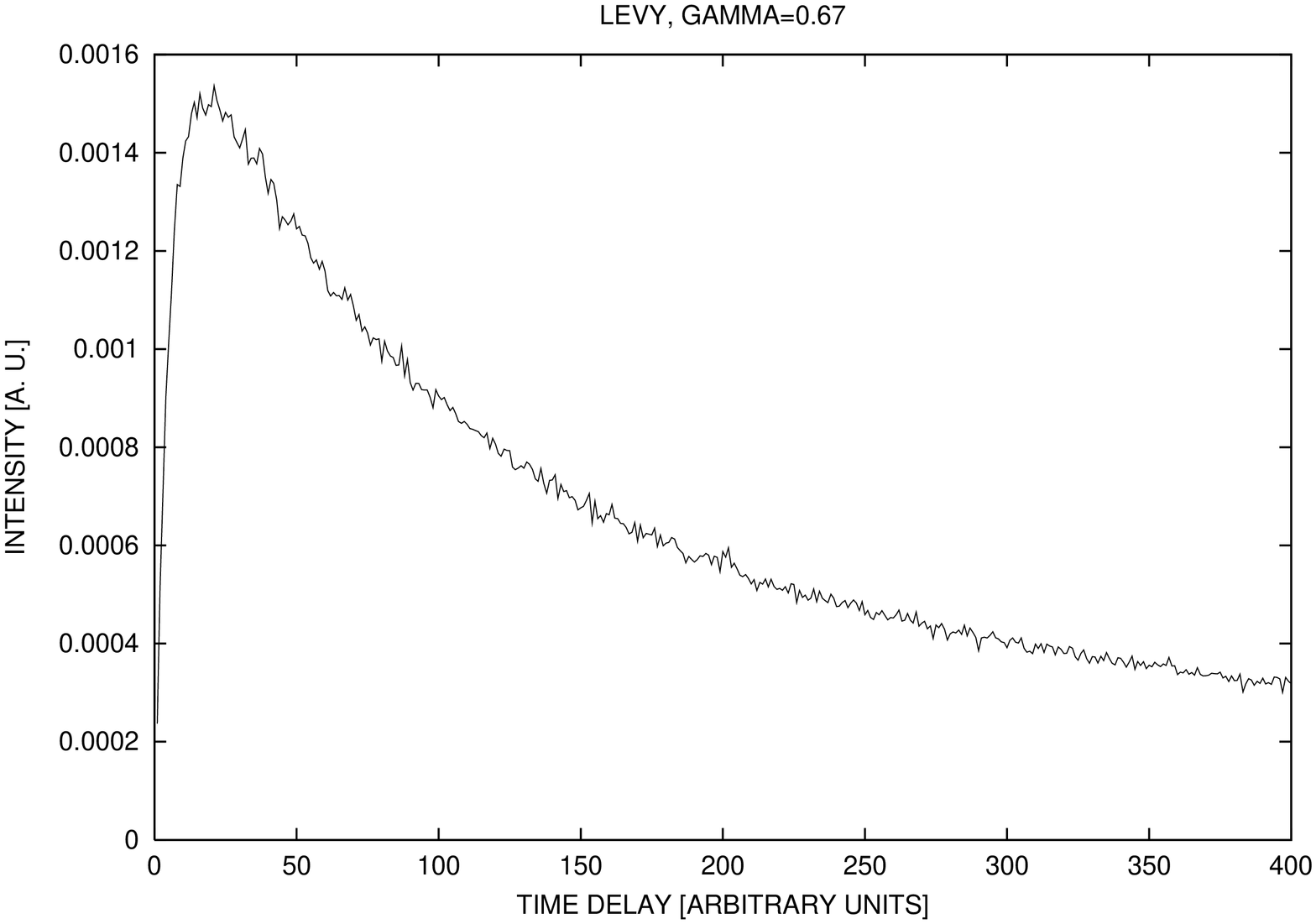,width=3.5in,angle=-90}}
\caption{Intensities of pulses predicted by the Gaussian model ($\gamma=2$),  
by the L\'evy-Cauchy model ($\gamma=1$), and by 
the L\'evy model with $\gamma=2/3$,   
for statistically uniform  
distribution of electron density is the interstellar medium. Note the 
sharp rise and the pointed apex of the signal predicted by the L\'evy 
model compared to the signal predicted by the Gaussian model. The shapes 
are rescaled to have similar decaying parts.}
\label{levy_cauchy}
\end{figure}
}

\newpage
{
\begin{figure} [tbp]
\centerline{\psfig{file=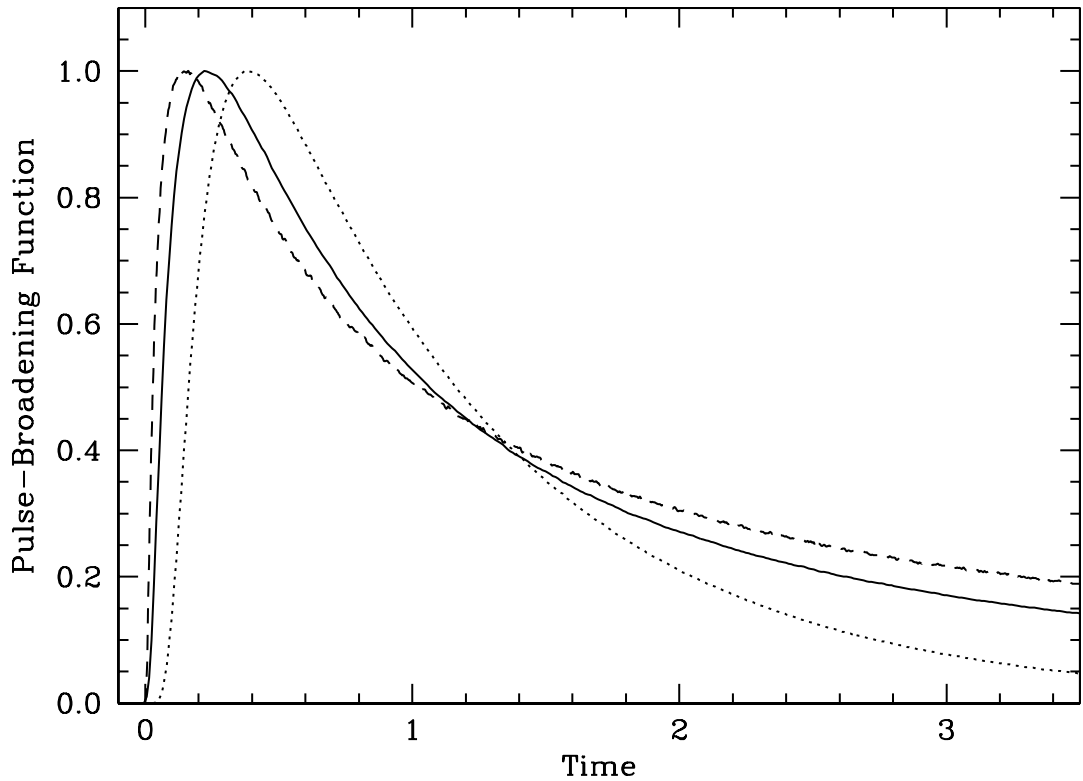,width=6.in,angle=0}}
\caption{
Pulse broadening function for $\gamma = 2/3$ (dashed line), $\gamma=1$
(Cauchy: solid line),
and $\gamma=2$ (Gaussian: dotted line). The unscattered signal would
arrive at time $t=0$.
Note the different relative arrival times of the peaks,
as well as the different power-law behavior in the tail, for different
values of $\gamma$. The shapes are rescaled to have same amplitudes 
and half-widths.}
\label{compare}
\end{figure}
}


\newpage
{
\begin{figure} [tbp]
\centerline{\psfig{file=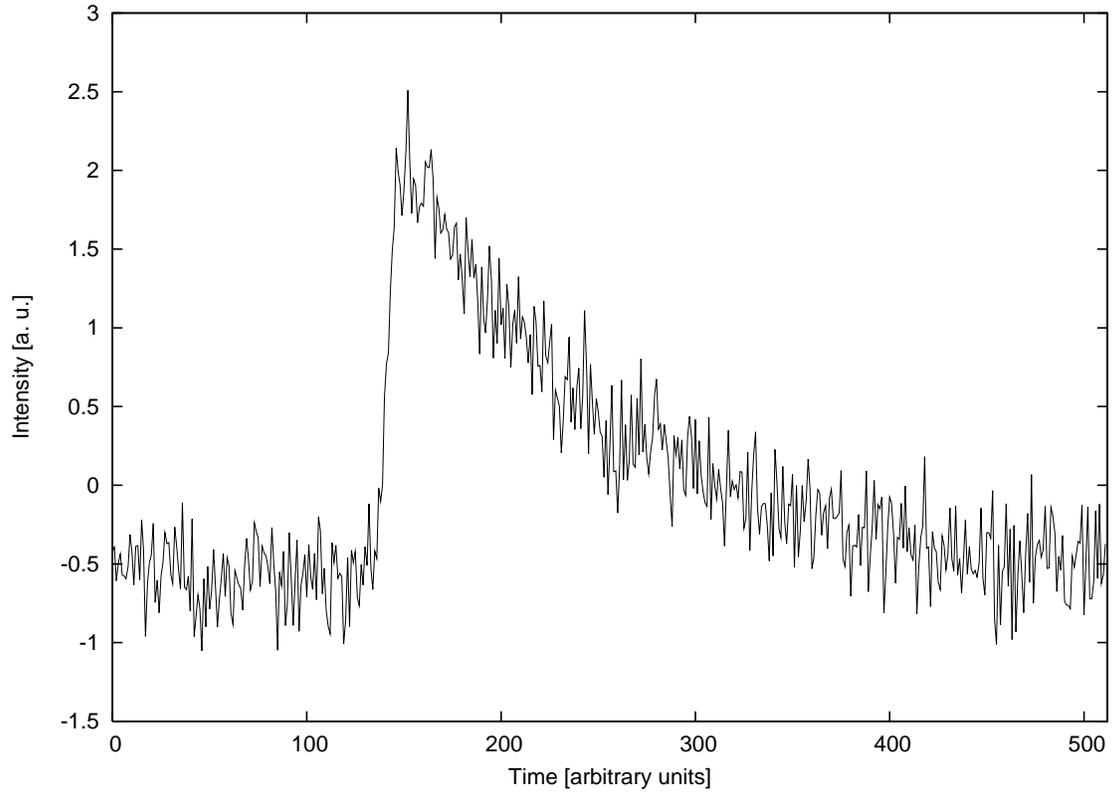,width=6.in,angle=-90}}
\vskip3mm
\caption{The pulse shape of the pulsar  
J1848-0123, obtained by 
Ramachandran, Mitra, Deshpande, Connell, \& Ables~(1997) (Courtesy of 
Ramachandran Rajagopalan). 
The data are obtained with the Ooty Radio telescope 
at 327 MHz. Note the sharp rise and the pointed apex of the signal.} 
\label{shapes2}
\end{figure}
}

\newpage

{
\begin{figure} [tbp]
\centerline{\psfig{file=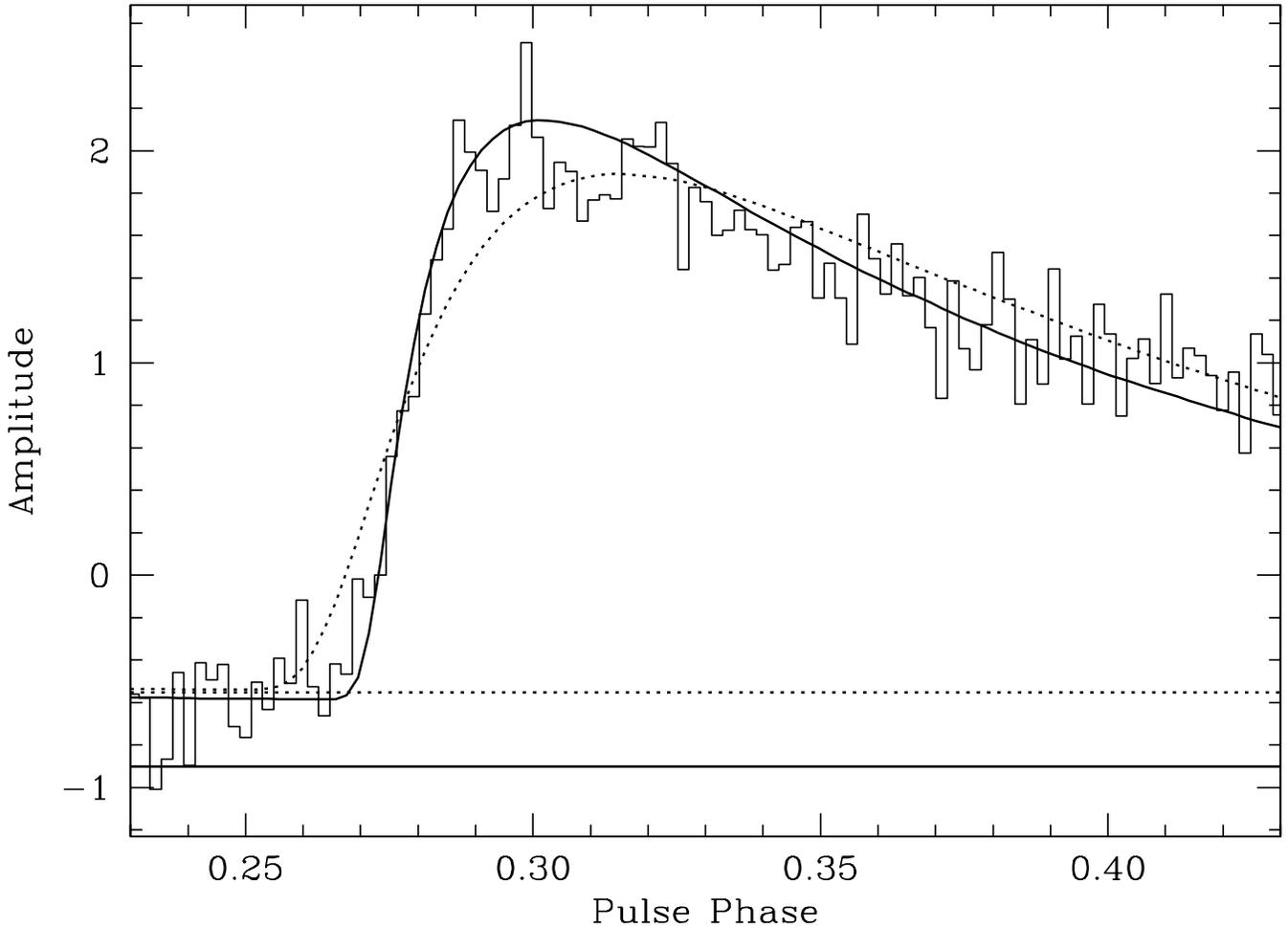,width=7.5in,angle=0}}
\caption{Best-fitting pulse-broadening functions for the pulse 
of pulsar J1848-0123. 
Solid line: $\gamma = 1$ (Cauchy), dotted line: $\gamma=2$ (Gaussian).
Histogram shows the pulse obtained by Ramachandran et al. (1997). [Courtesy of 
Ramachandran Rajagopalan.] 
Horizontal lines at bottom show the zero levels for the two functions; note
that the intensity never reaches zero for $\gamma=1$, as a consequence of
overscattering.
Also note that the Cauchy curve yields a steeper rise and a sharper peak,
in agreement with the data.
}
\label{alpack}
\end{figure}
}

\end{document}